\documentclass[%
reprint, 
prx,
amsmath,amssymb,
aps,
floatfix,
]{revtex4-2}

\usepackage{graphicx}
\usepackage{bm}
\usepackage[colorlinks, allcolors=blue]{hyperref}
\usepackage{amsmath}
\usepackage{physics}
\usepackage{booktabs}
\usepackage{caption}
\usepackage{subcaption}
\usepackage{bm}

\def\urbana{
The Anthony J. Leggett Institute for Condensed Matter Theory and IQUIST and NCSA Center for Artificial Intelligence Innovation and Department of Physics, University of Illinois at Urbana-Champaign, IL 61801, USA}

\begin{document}
\title{Neural network backflow for ab-initio quantum chemistry}

\author{An-Jun Liu} 
\affiliation{\urbana}
\author{Bryan K. Clark}
\affiliation{\urbana}

\begin{abstract}
The ground state of second-quantized quantum chemistry Hamiltonians provides access to an important set of chemical properties.  Wavefunctions based on ML architectures have shown promise in approximating these ground states in a variety of physical systems. In this work, we show how to achieve state-of-the-art energies for molecular Hamiltonians using the the neural network backflow wave-function.  To accomplish this, we optimize this ansatz with a variant of the deterministic optimization scheme based on SCI introduced by [Li, et. al JCTC (2023)] which we find works better than standard MCMC sampling.  For the molecules we studied,  NNBF gives lower energy states than both CCSD and other neural network quantum states.  We systematically explore the role of network size as well as optimization parameters in improving the energy.  We find that while the number of hidden layers and determinants play a minor role in improving the energy, there is significant improvements in the energy from increasing the number of hidden units as well as the batch size used in optimization with the batch size playing a more important role. 

\end{abstract}
\maketitle

\section{Introduction}
The solution to the electronic Schr\"{o}dinger equation for a given molecular system provides comprehensive access to its chemical properties from first principles. However, analytically solving the Schr\"{o}dinger equation is feasible only for hydrogen-like atoms in real space. Exact approaches in second quantization, such as the full configuration interaction (FCI) method, are intractable for large systems due to the exponential growth of the Hilbert space with the system size. The many-body electronic problem is fundamentally NP hard \cite{Whitfield2013,Gorman2022}, motivating efforts to develop numerical methods for approximating solutions in ab-initio quantum chemistry (QC).

Configuration interaction (CI) methods \cite{DavidSherrill1999} consider excitations above the Hartree-Fock (HF) reference state up to a fixed order, while coupled cluster (CC) approaches \cite{Coester1960} use exponential excitation operators up to a certain order to access all excited states. Although CC techniques are often considered the ``gold standard'' in ab-initio QC, they may fail in regimes of strong static correlations \cite{Bulik2015}.

Alternatively, variational wave-functions 
such as the Slater-Jastrow (SJ) \cite{Foulkes2001} and matrix-product states (MPS) \cite{White1992, White1999}, can be optimized to approximate the ground state.   Various approaches build on top of the SJ approach such as multi-determinant Jastrow wave-functions \cite{Clark2011} or  alternatively backflow transformation \cite{Feynman1956}, which renders orbitals dependent on the coordinates of all electrons  \cite{Kwon1993}.   Projector methods like diffusion Monte Carlo (DMC) \cite{Anderson1975}  systematically refine the variational starting point.   However, the effectiveness of these approaches is often bounded by the expressiveness of the wave function and the efficiency of the optimization.  For instance, MPS are inadequate for highly entangled systems and face challenges in dimensions greater than one.   Slater-Jastrow-Backflow (SJB) are often hard to systematically improve. 

Machine learning architectures presents significant promise as concise wave-functions to address certain constraints of existing variational representations.  Ref.~\onlinecite{Carleo2017} initially demonstrated that restricted Boltzmann machines (RBM), trained via variational Monte Carlo (VMC), could effectively capture the intricate correlations in certain interacting spin models and numerous advances  \cite{Choo2019, Choo2018, Nagy2019, Sharir2020}  since then have improved neural network quantum states (NNQS) for spin models.  More recently, there has been significant work using machine learning architectures to develop wave-functions for Fermion systems \cite{Nomura2017,Francesco2019,Stokes2020,Lin2023,Zhuo2022,Di2019,zejun2023,Pfau2020,Hermann2020,Choo2020,Barrett2022,Zhao2023,Li2023,Shang2023,Wu2023,Malyshev2023}.  

The advent of wave-functions for Fermions has motivated an interest in addressing molecular and QC Hamiltonians.  The FermiNet \cite{Pfau2020} and PauliNet \cite{Hermann2020} wave-functions, which are real-space generalizations of the Neural Net Backflow (NNBF) \cite{Di2019}, have shown to be competitive with other ab-intio QC methods in continuous space.  Using a  second quantized Hamiltonian in a finite basis,  ref.~\onlinecite{Choo2020} mapped the molecular system onto an equivalent spin problem and demonstrated that RBM could achieve 
higher accuracy than CCSD(T).  
However, they noted that using ordinary Markov chain Monte Carlo (MCMC) methods for sampling was inefficient due to sharp peaks in the underlying probability distribution around the HF state and neighboring excited states.  To address this sampling inefficiency as well as gain more accuracy than the RBM, subsequent studies \cite{Barrett2022, Zhao2023, Shang2023, Wu2023} introduced autoregressive neural networks (ARN) for exact sampling. Further improvements in optimization were introduced by  ref.~\onlinecite{Li2023} which proposed a non-stochastic optimization scheme adapted from the selected configuration interaction (SCI) method to deterministically select samples on-the-fly leading to competitive molecular energies using RBMs. Interestingly, despite NNBF style ansatz demonstrating high accuracy in the continuum for QC, the NNBF approach has not yet been applied to second quantized QC Hamiltonians. 

In this work, we address this omission using the NNBF \cite{Di2019}  to approximate the ground state of molecular systems in second quantization.  To optimize our wave-function, we apply and demonstrate the superiority of a deterministic fixed-size selected configuration (FSSC) iteration scheme,  adapted from ref.~\onlinecite{Li2023}.  Our experiments on various molecules indicate that the NNBF ansatz not only surpasses traditional CCSD benchmarks but also finds lower energies than all other existing NNQS methods on large molecules (and equally good energies on smaller molecules).  We also evaluate the dissociation curve for diatomic Nitrogen demonstrating NNBF's ability to capture both strong and weak quantum correlation. Additionally, we conduct a systematic investigation into the impact of network architecture and batch size, revealing that increasing the batch size is a more efficient and effective approach to improving NNBF performance.

\section{Methods}
In this section, we provide a detailed description of the structure of NNBF and the workflow of FSSC.
\subsection{Neural Network BackFlow Architecture}
For a molecular system containing $N_e$ electrons, a basis set composed of $N_o (> N_e)$ spin-orbitals $\mathcal{B}=\{ \ket{\phi_i} \}_{i=1}^{N_o}$ is given to define a many electron wave-function in the form of $\ket{\psi} = \sum_{i=1}\psi(\bm{x_i})\ket{\bm{x_i}}$ where $\ket{\bm{x_i}}=\ket{x_i^1,\dots,x_i^{N_o}}$ is the i-th computational basis vector in second quantization. Here, $x_i^j \in \{0,1\}$ denotes whether the j-th spin-orbital is occupied in the i-th computational basis vector.

\begin{figure}[t]
    \centering    
    \includegraphics[width=\linewidth]{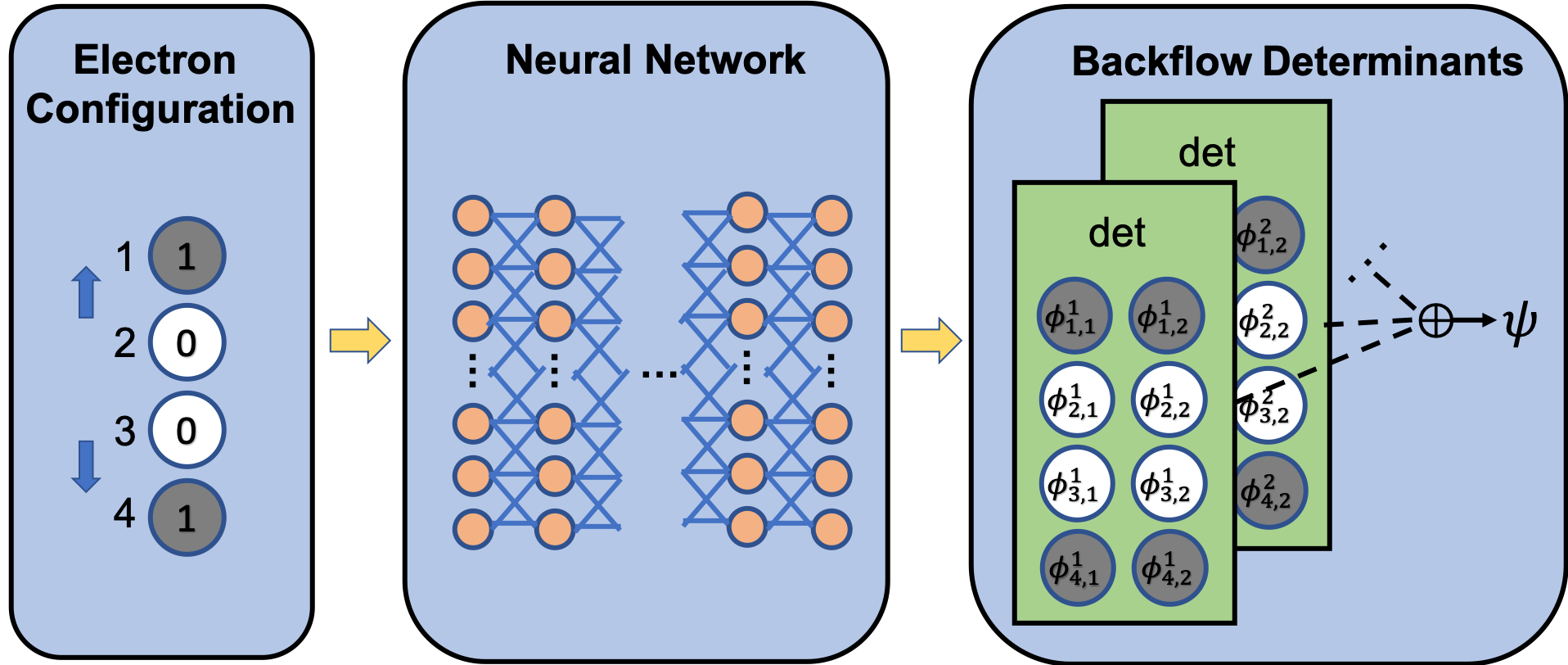}  
    \caption{Illustration of the NNBF architecture with an example input configuration: 2 spin-orbitals with 1 spin-up electron occupying the first spin-orbital and 1 spin-down electron occupying the second spin-orbital. The neural network takes the configuration string as input and outputs a set of $D$ configuration-dependent single-particle orbitals of shape $(N_o,N_e)$. Rows of these orbitals are then selected based on the electron locations to form square matrices, from which determinants are computed. For this example, the first and last rows (the gray orbitals) are selected to compute the determinant. The sum of these determinants yields the amplitude for the input configuration.}    
    \label{fig:NNBF_structure}
\end{figure}

The NNBF consists of a multilayer perceptron (MLP) which generates the "configuration-dependent" single particle orbitals
used in evaluating the Slater determinant (see Figure \ref{fig:NNBF_structure}).
The MLP directly takes the configuration strings as input feature vector and feeds it through a series of intermediate layers consisting of a fully-connected feed-forward neural network (FNN) combined with a ReLU nonlinear activation function i.e., $\textbf{h}^{l+1}=ReLU(\textbf{W}^l \textbf{h}^{l} + \textbf{b}^{l})$. Instead of directly returning a scalar wave-function amplitude, the MLP outputs a configuration-dependent additive change to the set of single-particle orbitals for each determinant. In practice, the final output vector of the network, $\textbf{y}=\textbf{W}^L \textbf{h}^{L} + \textbf{b}^{L}$, will be reshaped into an array of $D$ matrices, where each matrix $\phi^k_{ij}$ is of shape $(N_o, N_e)$. 
The i-th row of the k-th matrix has a physical interpretation that it is the evaluation of the k-th set of configuration-dependent orbitals if the i-th spin-orbital is occupied.
A square matrix can be generated by extracting the rows whose indices correspond to the locations of the electrons in the input configuration, and the amplitude of this configuration is defined as $\psi_{\theta}(\bm{x})=\sum_{k=1}^{D}det[\phi^k_{i=\{l | x_l=1\},j}]$ where $\theta$ is the model parameters. Consequently, the full NNBF wave-function is described as $\ket{\psi_{\theta}} = \sum_{i=1}\psi_{\theta}(\bm{x_i})\ket{\bm{x_i}}$.

\subsection{Wavefunction Optimization}
Given a quantum chemical Hamiltonian, 
\begin{equation}
    \hat{H} = \sum_{i,j,\sigma}t_{ij}c_{i,\sigma}^\dagger c_{j,\sigma} + \frac{1}{2} \sum_{i,j,k,l,\sigma,\sigma'}V_{ijkl}c_{i,\sigma}^\dagger  c_{j,\sigma'}^\dagger c_{l,\sigma'} c_{k,\sigma}.
\end{equation}  
our target is to minimize the variational energy of our wave-function, 
    \begin{equation} \label{eq:exact_energy}        E_\theta=\frac{\bra{\psi_\theta}\hat{H}\ket{\psi_\theta}}{\braket{\psi_\theta}{\psi_\theta}}=\mathbb{E}_{p_{\theta}(\bm{x})} \left[E_{loc}(\bm{x})\right]
\end{equation}
which can be evaluated by MCMC by sampling from the probability distribution $p_\theta(x) \propto |\Psi_\theta(x)|^2$  where the   local energy $E_{loc}(\bm{x}) = \frac{\bra{\psi_\theta}\hat{H}\ket{\bm{x}}}{\bra{\psi_\theta}\ket{\bm{x}}}$.     To optimize this energy, we will estimate the gradient of it, $\nabla_\theta E_\theta$,  and update $\theta$ using ADAM \cite{Adam}.  
We perform a CISD pretraining before optimizing with expected energy; the implementation details of this pretraining are provided in Appendix \ref{sec:CISD_pretraining}.

One approach for computing the gradient
\begin{equation} \label{eq:exact_gradient}
    \nabla_\theta E_\theta = 2\Re{\mathbb{E}_{p_{\theta}(\bm{x})}[\left[E_{loc}(\bm{x}) - E_\theta \right] \nabla_\theta \ln{\abs{ \psi_\theta(\bm{x})}}]}
\end{equation}
is through an MCMC sampling of the energy.  For optimizing molecular Hamiltonians, we find that this is a suboptimal way to update the variational parameters.  A similar observation was made by ref.~\onlinecite{Choo2020} which found that typical MCMC sampling technique is inefficient for the ground state of molecular system due to the peaks around the HF state and neighboring excited states \cite{Bytautas2009,Anderson2018} which results in drawing many identical samples from the dominant states when sampling.  Various techniques have been developed to overcome these issues. 
For autoregressive variational wave-functions, the probability of a configuration $\bm{x}$ can be modeled using its conditional probabilities, expressed as $p_\theta(\bm{x})=\prod_i p_\theta(x_i | x_1,x_2,\dots,x_{i-1})$. This autoregressive property, combined with more efficient techniques, enables sampling the numbers of occurrences for each unique string, thus avoiding duplicate sampling efforts \cite{Barrett2022,Zhao2023,Shang2023,Malyshev2023}. Recently, Li et al. \cite{Li2023} proposed an algorithm based on SCI to select a set of states and approximate their frequency of occurrences instead of statistically sampling them.
This set of states is a deterministic selection of a set of important configurations identified through the modulus of the wave-function ansatz during energy evaluation.

\subsection{Deterministically Select Sample Space}
\label{sec:fssc_implemenation}
\begin{figure}[t]
    \centering    \includegraphics[width=\linewidth]{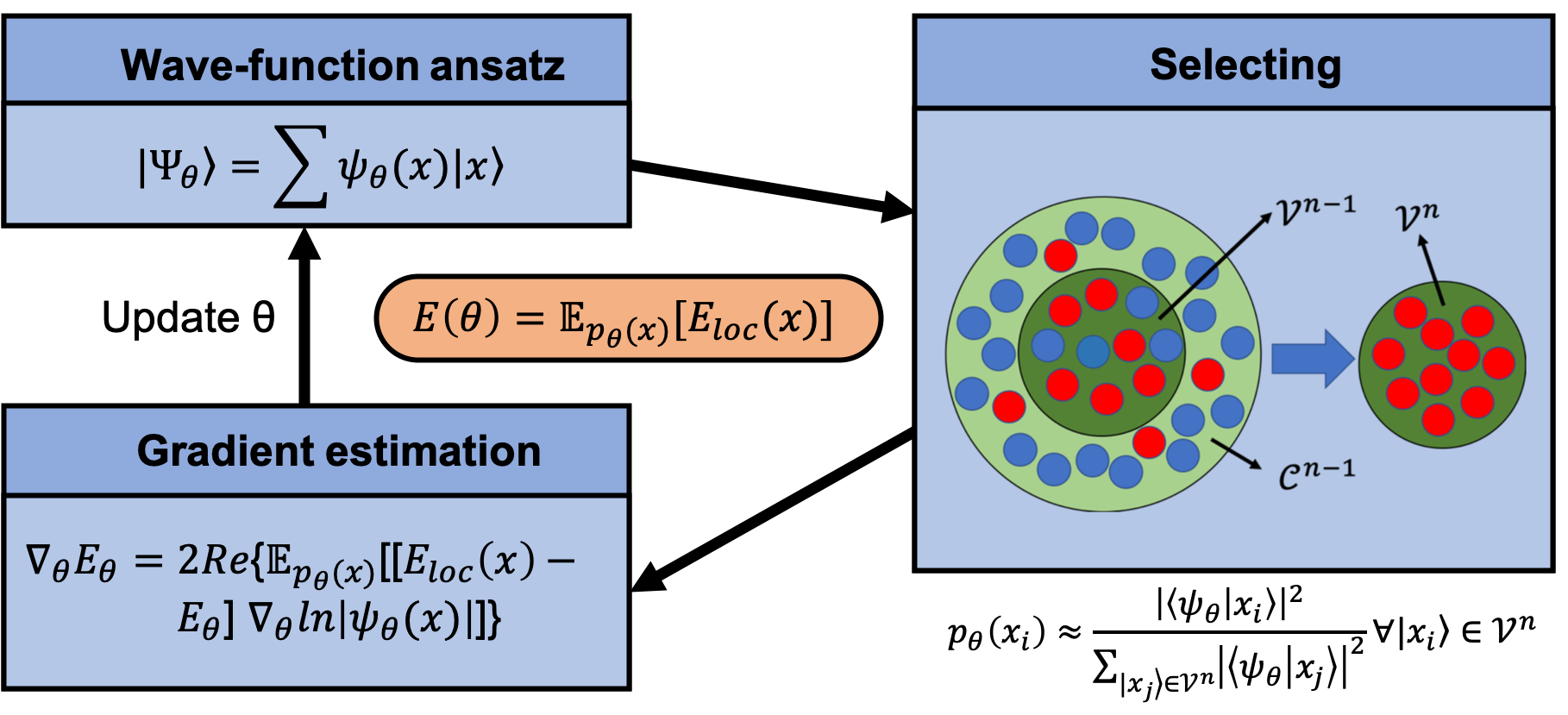}  
    \caption{A diagrammatic description of the workflow of FSSC. Each circle represents one configuration. Initially, the algorithm initializes a parametrized wave-function and a core space $\mathcal{V}^{0}$ of size $N_u=10$. After each iteration, the amplitude moduli for all configurations in $\mathcal{V}^{n-1} \cup \mathcal{C}^{n-1}$ are computed, and the 10 largest unique ones (denoted by red configurations) are selected to form the new core space $\mathcal{V}^{n}$. The energy (depicted as the loss function in the orange box) and its gradient are estimated by constraining the relative sum to only consider $\mathcal{V}^{n}$, and the latter is used to update the model parameters.}    
    \label{fig:FSSC_workflow}
\end{figure}

In this work, we will use a variant of the scheme of ref.~\onlinecite{Li2023} which we call the FSSC scheme. For completeness, we describe the entire FSSC approach below while a diagrammatic description of it is also provided in Figure \ref{fig:FSSC_workflow}. 

The full configuration space is divided into three parts as in a typical SCI method: the core space $\mathcal{V}$ consisted of the dynamically selected important configurations, the connected space $\mathcal{C}$ containing all other configurations connected to the current core space through nonzero Hamiltonian matrix elements, and the remaining part of the configuration space. The HF state and some neighboring excited states are deterministically pre-selected to form the initial core space $\mathcal{V}^0$ of size $N_u$. After each iteration of optimization,  we find the connected space  $\mathcal{C}^{n-1}$ induced from the core space $\mathcal{V}^{n-1}$ from the last iteration.  The new core space  $\mathcal{V}^{n}$ will be deterministically chosen by selecting the $N_u$ configurations from the union of $\mathcal{C}^{n-1}$ and $\mathcal{V}^{n-1}$ which have the largest unique amplitudes with respect to the new state with updated parameters. 
This $\mathcal{V}^{n}$ will then be used to evaluate an approximation to the gradient and update the model parameters: 
\begin{align}\label{eq:fssc_grad}
    \nabla_\theta E_\theta = 2\Re\left\{\sum\nolimits_{\ket{\bm{x_i}}\in \mathcal{V}^{n}} p_\theta(\bm{x_i}) 
    \left[\left[ E_{loc}(\bm{x_i}) \right.\right.\right. \nonumber\\
    \left.\left.\left.- E_\theta \right] \nabla_\theta \ln{\abs{ \psi_\theta(\bm{x_i})}}\right] \right\}
\end{align}
where 
\begin{equation}\label{eq:FSSC_energy}
E_\theta = \sum\nolimits_{\ket{\bm{x_i}}\in \mathcal{V}^{n}} p_\theta(\bm{x_i}) E_{loc}(\bm{x_i}).
\end{equation} 
and 
\begin{equation}\label{eq:fussy_prob}
    p_\theta(\bm{x_i}) = \frac{\lvert \braket{\psi_\theta}{\bm{x_i}}\rvert^2}{\sum_{\ket{\bm{x_j}}\in \mathcal{V}^{n}}\lvert \braket{\psi_\theta}{\bm{x_j}}\rvert^2}
\end{equation} 
where the relevant sums are over the core space. Approximating the various quantities above by summing over only the core space  performs well when the ground state is predominantly influenced by a small number of significant configurations, which is a typical characteristic of molecular systems in the canonical HF basis.  We note that all the energies reported in this work (not including traces over optimization time) are computed exactly by Monte Carlo (MC) sampling.  See appendix \ref{sec:experimental_setup} for details. 

The major distinction between our approach and ref.~\onlinecite{Li2023} lies in fixing the size of the core space, as opposed to selecting configurations whose modulus relative to the largest one is greater than a selection cutoff, $\epsilon$. This allows us to leverage the advantages of just-in-time (JIT) compilation with JAX, as the shape of the input array to the jitted functions remains constant throughout iterations. Additionally, a fixed number of unique configurations determines the computational cost, which is directly proportional to $N_u$, whereas in most other NNQS works \cite{Barrett2022, Li2023, Choo2020, Shang2023}, only the total number of samples could be specified. Moreover, a fixed batch size prevents the possibility of an unexpectedly large number of configurations overwhelming the limited computational resources, whereas the core space could potentially encompass the entire configuration space at some point in the ref.~\onlinecite{Li2023}. At one point, we have compared a fixed batch size to a dynamically adjusted batch size. The optimized energy using a selection cutoff is essentially the same as the optimized energy with a fixed batch size which is equal to the average batch size of the selection cutoff (Fig. \ref{fig:NN_bs_Li2O}).

\section{Results}

Here we showcase the effectiveness of the FSSC scheme by comparing it to the typical MCMC scheme, evaluate the performance of the NNBF on various molecules, train the NNBF model to emulate the dissociation curve of nitrogen molecule, and investigate the impact of network architecture and batch size on NNBF. Further details on the default network architectures, hyperparameters, training routine, and the post-training MCMC inference routine can be found in Appendix \ref{sec:experimental_setup}. 

\subsection{FSSC vs. Traditional MCMC Optimization}

To demonstrate the efficacy of the FSSC scheme in addressing the inefficiencies of MC sampling, we conduct a comparison on lithium oxide using both the FSSC scheme and the standard MCMC scheme.  For a fair comparison, we set the number of unique configurations in the FSSC scheme equal to the number of walkers in the MCMC scheme, i.e. $N_u=N_w=8192$.

\begin{figure}[]
    \centering
    \includegraphics[width=\linewidth]{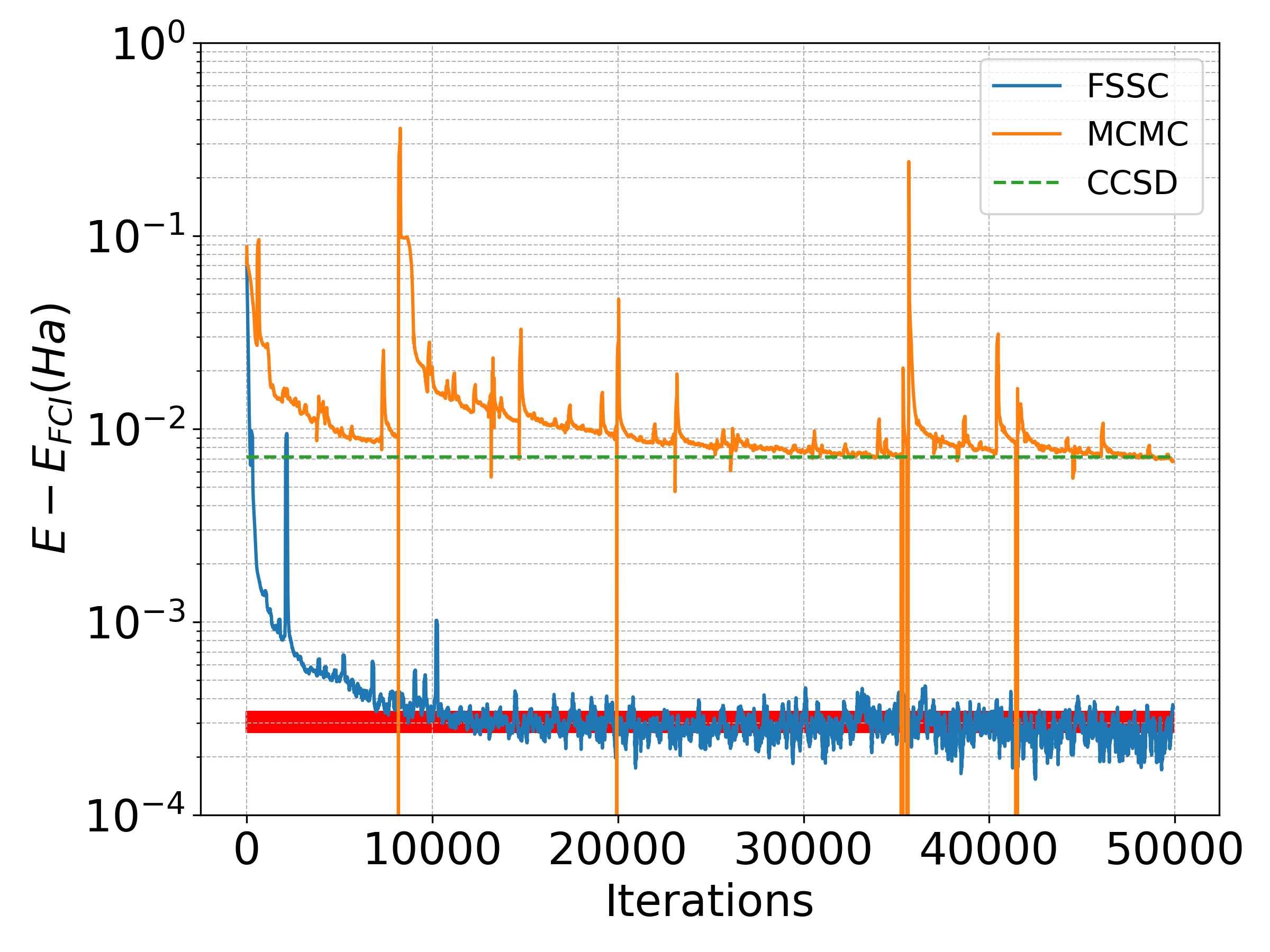}
    \caption{Comparison between FSSC and MCMC schemes on lithium oxide ($N_o=30$, $N_e=14$, $N_u=N_w=8192$). The red band represents the post-training MCMC inference energy for the FSSC scheme with a width of $\sigma$  in each direction around the mean. A moving average window of 100 is applied to improve readability.}
    \label{fig:MCMC_vs_FSSC}
\end{figure}

In Figure \ref{fig:MCMC_vs_FSSC} we see both faster convergence as well as a lower converged energy for the  FSSC scheme compared to the conventional MCMC scheme, which happens to only converge to the CCSD baseline.  Additionally, we observe that there are only 111 unique configurations in the 8192 MCMC samples at convergence, with the most frequent sample being the HF configuration appearing in 7686 walkers. This observation validates the inefficient MC sampling discovered by ref.~\onlinecite{Choo2020} and highlights the FSSC scheme's ability to effectively capture important configurations in large configuration space, akin to conducting MC sampling with a sample size $N_w \gg N_u$. Additionally, the FSSC method shows a smoother, more monotonic trend in energy values during the optimization steps, especially compared to the MCMC scheme, which, although it displays an average reduction in energy, does so with noticeable oscillations. This demonstration aligns with the results presented by ref.~\onlinecite{Li2023}. 

\subsection{Benchmark With Other QC Methods} \label{sec:GSE}

We assess the performance of our algorithm by comparing molecular ground state energies with CCSD baselines and other NNQS results, utilizing a list of molecules with geometries sourced from \textit{PubChem} \cite{pubchem}. Table \ref{tab:GSE} demonstrates that  NNBF not only surpasses conventional CCSD methods but also is equal or lower in energy than all other existing NNQS approaches, particularly showing superiority in larger molecular systems.

\begin{figure}[t]
    \centering
    \includegraphics[width=\linewidth]{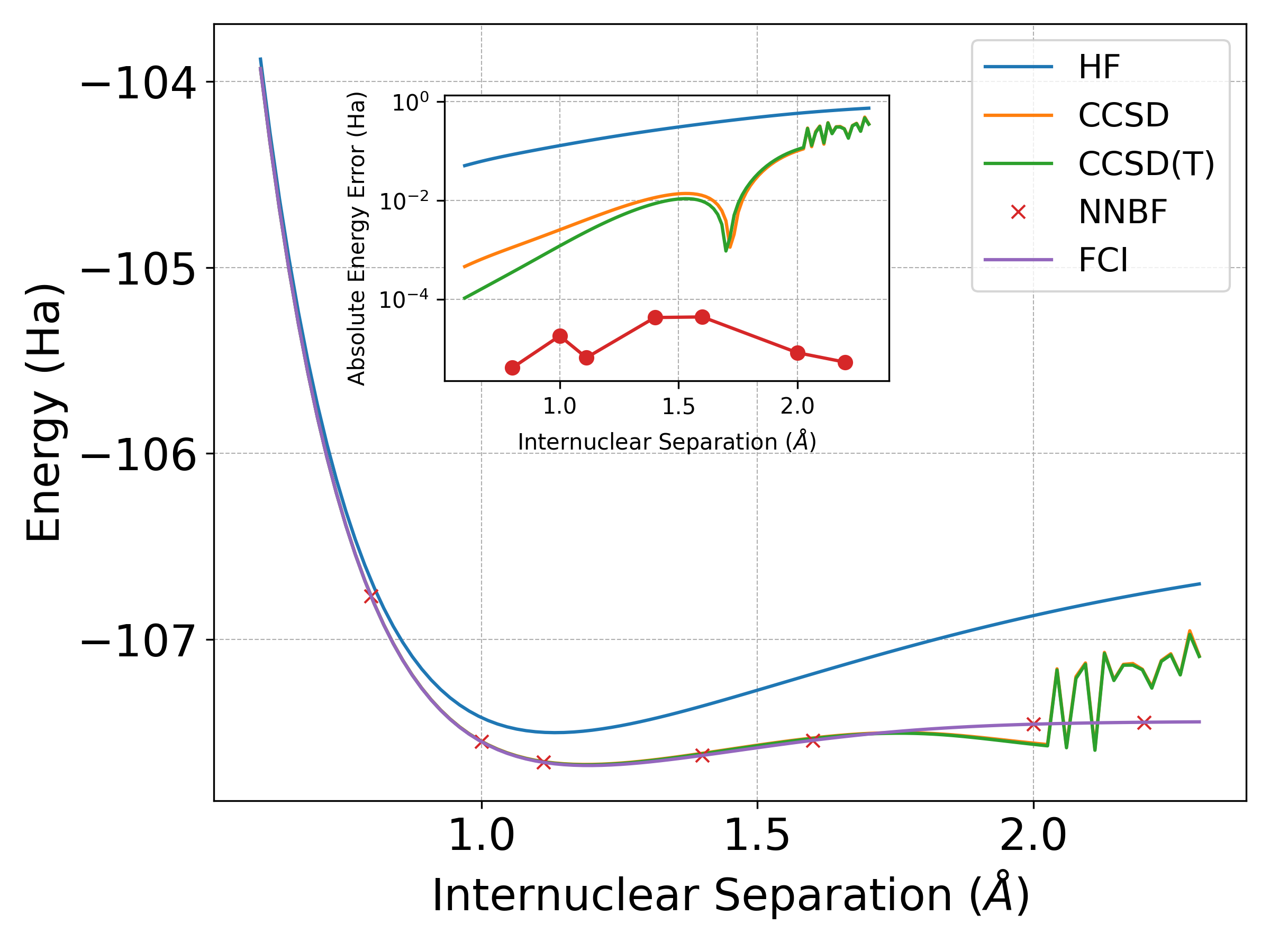}
    \caption{Dissociation curve for N$_2$ obtained with NNBF, HF, CCSD, and CCSD(T) methods. The FCI energy is used as the ground-truth energy. The NNBF energy is trained using the FSSC scheme with $N_u=4096<N_t$, and the reported energy here is computed exactly, as it remains feasible to compute.}
    \label{fig:N2_curve}
\end{figure}

Additionally, we explore the NNBF's ability to capture quantum correlation by computing the dissociation curve of the diatomic nitrogen molecule. As illustrated in Figure \ref{fig:N2_curve}, the NNBF precisely matches the FCI energy and outperforms all other conventional QC approaches in scenarios involving both near-equilibrium geometry and bond breaking, where other methods falter due to the presence of strong quantum correlations at large bond separation. This experiment highlights   NNBF's capability to accurately capture both strong and weak correlations.

\begin{table*}[t] 
  \centering
  
  \begin{tabular}{ccccccc}
    \toprule
    Molecule & $N_t$ & $N_u$ & CCSD & FCI & Best NNQS & NNBF\\ 
    \midrule
    N$_2$ & 14400 & 4096 & -107.656080 &-107.660206 & -107.6602$^a$ & -107.660198(5) \\
    CH$_4$ & 15876 & 4096 & -39.806022 & -39.806259 & -39.8062$^a$ & -39.806255(3)$^*$ \\
    LiF & 44100 & 4096 & -105.159235 & -105.166172 & -105.1661$^a$ & -105.166178(9)\\
    LiCl & 1002001 & 8192 & -460.847579 & -460.849618 & -460.8496$^b$ & -460.849615(17)\\
    Li$_2$O & 41409225 & 65536 & -87.885514 & -87.892693 & -87.8922$^a$ & -87.892645(9)\\
    C$_2$H$_4$O & 2538950544 & 131072 &-151.120474 & - & -151.120486$^c$ & -151.121703(68)\\
    \bottomrule
  \end{tabular}
  
  \caption{Ground state energies obtained by NNBF. The conventional CCSD and FCI results are listed for comparison. $N_t$ indicates the total number of physically valid configurations conserving the particle number and the total charge. Best NNQS column depicts best published NNQS energies, and footnotes mark the methods: a: QiankunNet \cite{Shang2023}, b: NAQS \cite{Barrett2022}, c: MADE \cite{Zhao2023}. The NNBF energy column displays the post-training MCMC inference energy. The energy marked by $^*$ is obtained by setting $D=3$.} 
  \label{tab:GSE}
\end{table*}

\subsection{The Effect of Network Architecture With Full Hilbert Space}\label{sec:NN_arch_full}

One of the critical factors influencing the performance of NNBF is the expressiveness of the ansatz, which is directly linked to the network architecture, including the number of hidden units $h$, the number of backflow determinants $D$, and the number of hidden layers $L$.  We first explore the impact of the size and shape of the network by comparing NNBF models with different values of $h$, $L$, and $D$ on methane. It's worth noting that we employ the entire Hilbert space ($N_u=N_t$) during optimization to eliminate the approximation error in equation~\eqref{eq:fssc_grad} and ~\eqref{eq:FSSC_energy} for small systems like methane. While this approach inherently encompasses the combinatorial complexity in the summation, making it non-scalable, it still provides insight into how the network architecture influences the representability of the ansatz and its performance.

\begin{figure*} 
  \centering

  \begin{subfigure}{0.32\linewidth}
    \includegraphics[width=\linewidth]{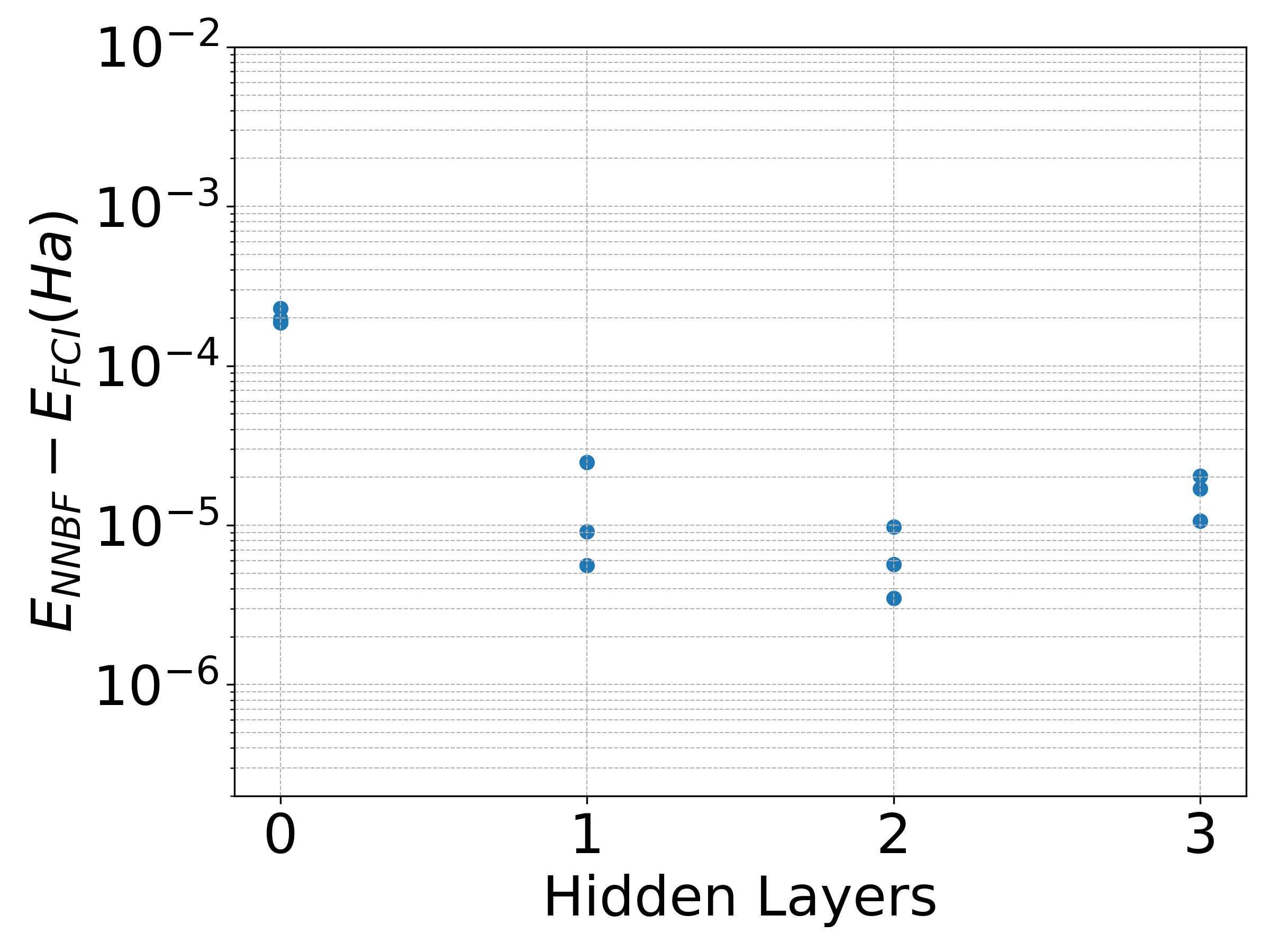}
    \caption{}
    \label{fig:NN_arch_full_width}
  \end{subfigure}
  \hfill
  \begin{subfigure}{0.32\linewidth}
    \includegraphics[width=\linewidth]{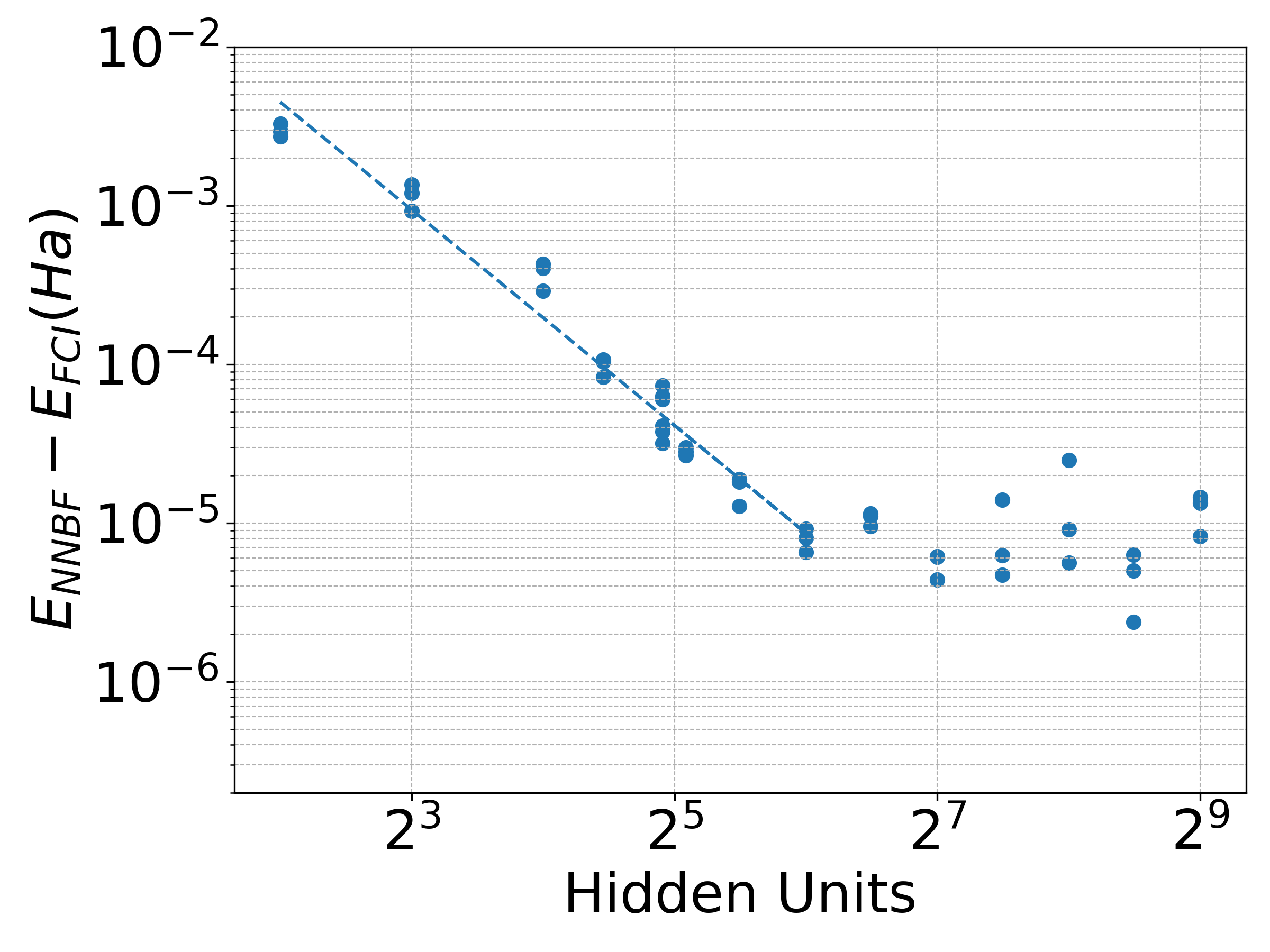}
    \caption{}
  \end{subfigure}
  \hfill
  \begin{subfigure}{0.32\linewidth}
    \includegraphics[width=\linewidth]{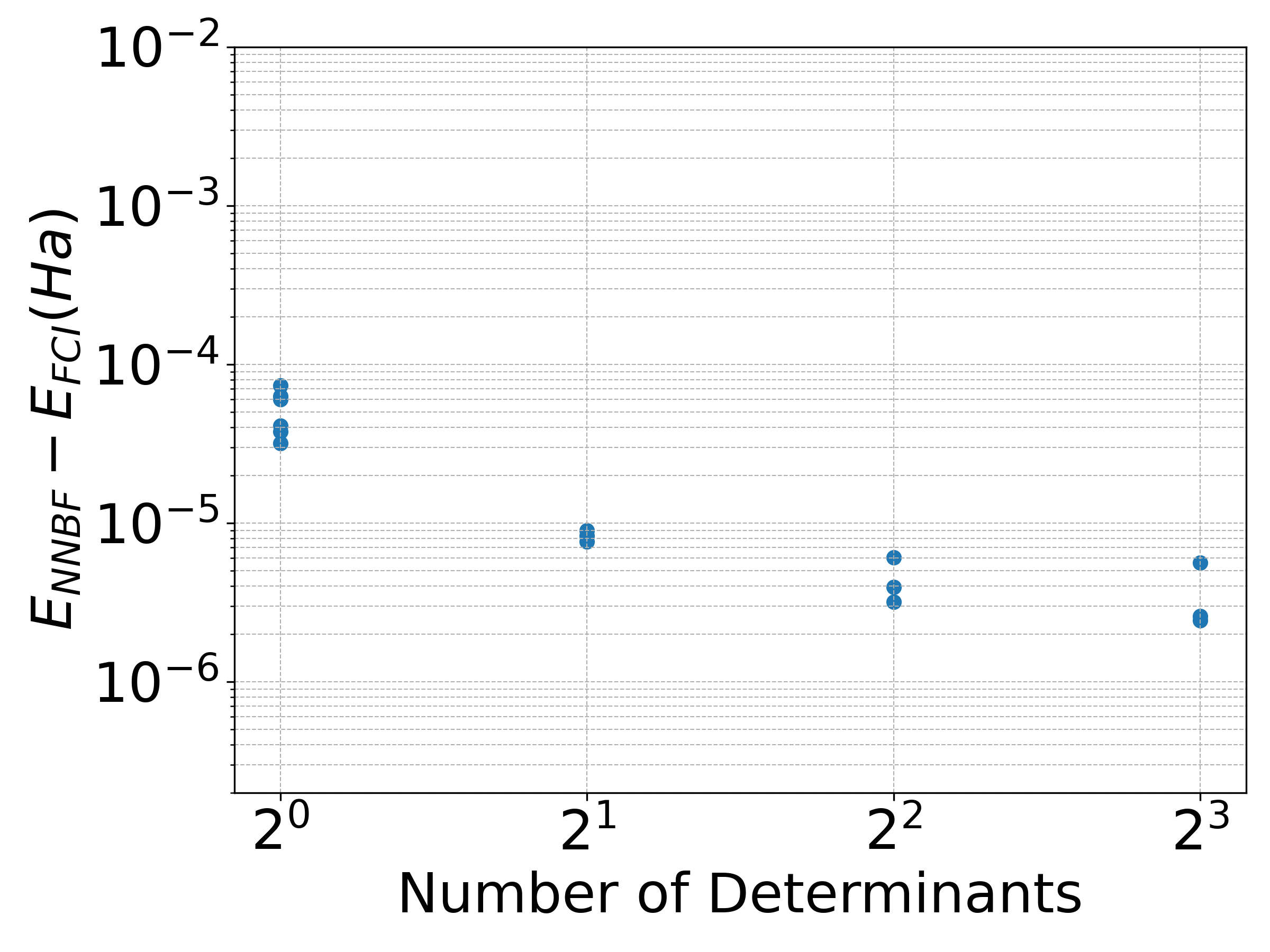}
    \caption{}
    \label{fig:NN_arch_full_det}
  \end{subfigure}
\caption{Effects of network architecture on the NNBF performance on CH$_4$ ($N_o=18$, $N_e=10$, $N_u=N_t$, $L=1$). Each point is one run of the same model. Double precision is employed for calculating the exact inference energy, as higher precision is necessary when the NNBF state closely approaches the true ground state. (a) Effect of network depth. The increase in performance levels off after the addition of 2 hidden layers to NNBF states. (b) Effect of number of hidden units. A wider hidden layer consistently improves accuracy, with the energy error decreasing at a rate of $\mathcal{O}(h^{-2.256228})$ with $R^2=0.968270$ until $h=64$. (c) Effect of number of determinants ($h=30$). Expanding the number of determinants does reduce the energy error and begins to plateau after 4 determinants.}
\label{fig:NN_arch_full}
\end{figure*}

As shown in Figure \ref{fig:NN_arch_full}, while there is an overall improvement in accuracy with an increase in the number of layers, the transition from 2 to 3 layers does not exhibit a significant enhancement; in fact, the energy increases slightly which suggets it gets caught in a slightly higher local minima. Additionally, increasing the number of hidden units results in a uniform improvement in accuracy before reaching saturation. According to a linear regression analysis of the log-errors, the error scales with the number of hidden units as $\mathcal{O}(h^{-2.253083})$. Moreover, augmenting the number of determinants reduces the energy error while keeping $L$ and $h$ fixed, although the benefits seem to plateau after reaching 4 determinants. This suggests that extensive linear combinations of determinants may not be essential as in multi-determinant SJ, consistent with findings reported in FermiNet \cite{Pfau2020}.

It's noteworthy that all curves in Figure \ref{fig:NN_arch_full} converge to an energy level close to the FCI energy within single precision. This indicates that these saturation points likely represent the global minima, as they have reached machine precision.

\subsection{The Effect of Network Architecture With Limited Batch Size}\label{sec:NN_arch_limited}

\begin{figure*} 
  \centering

  \begin{subfigure}{0.32\linewidth}
    \includegraphics[width=\linewidth]{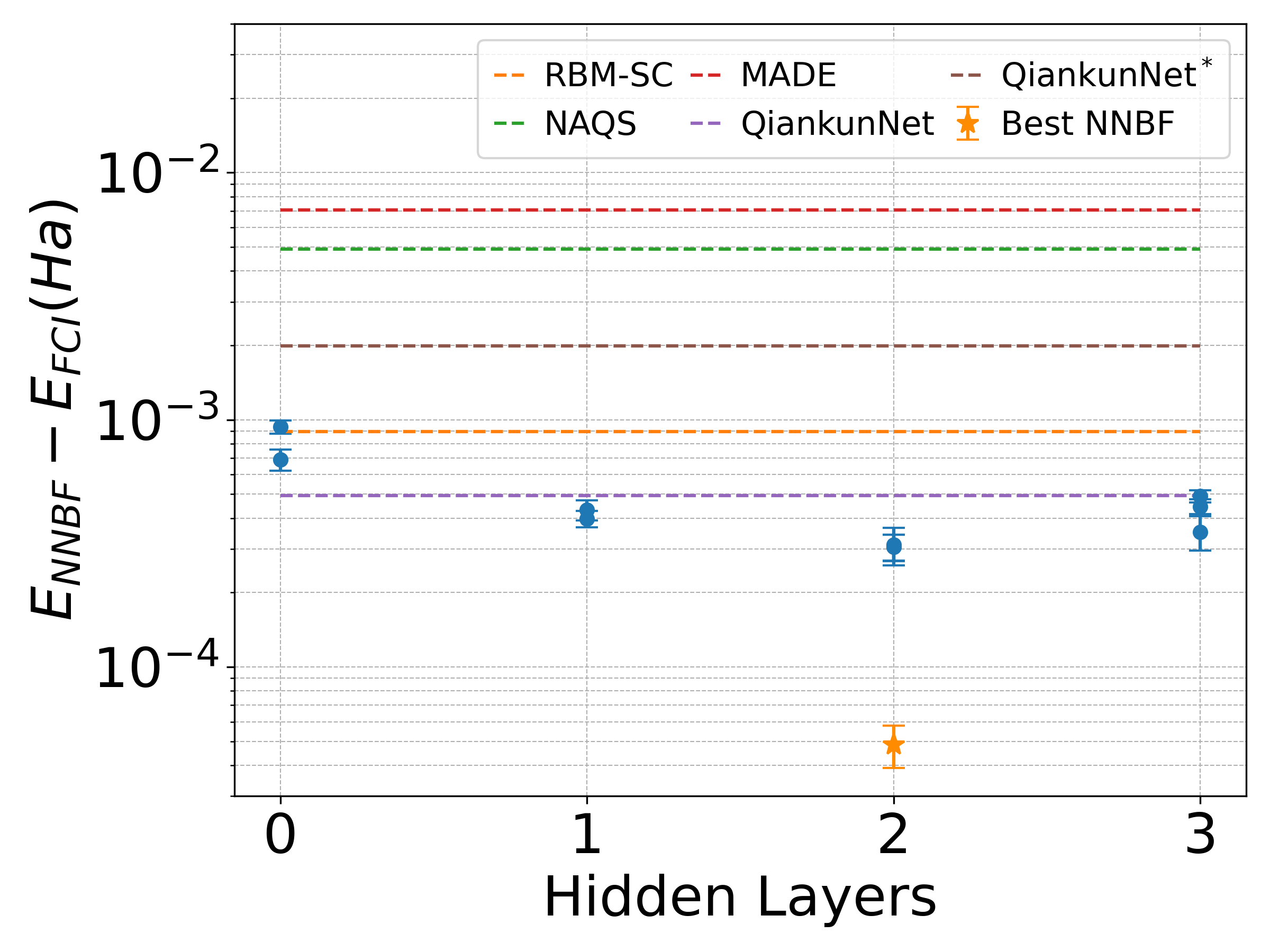}
    \caption{}
    \label{fig:NN_arch_limited_layer}
  \end{subfigure}
  \hfill
  \begin{subfigure}{0.32\linewidth}
    \includegraphics[width=\linewidth]{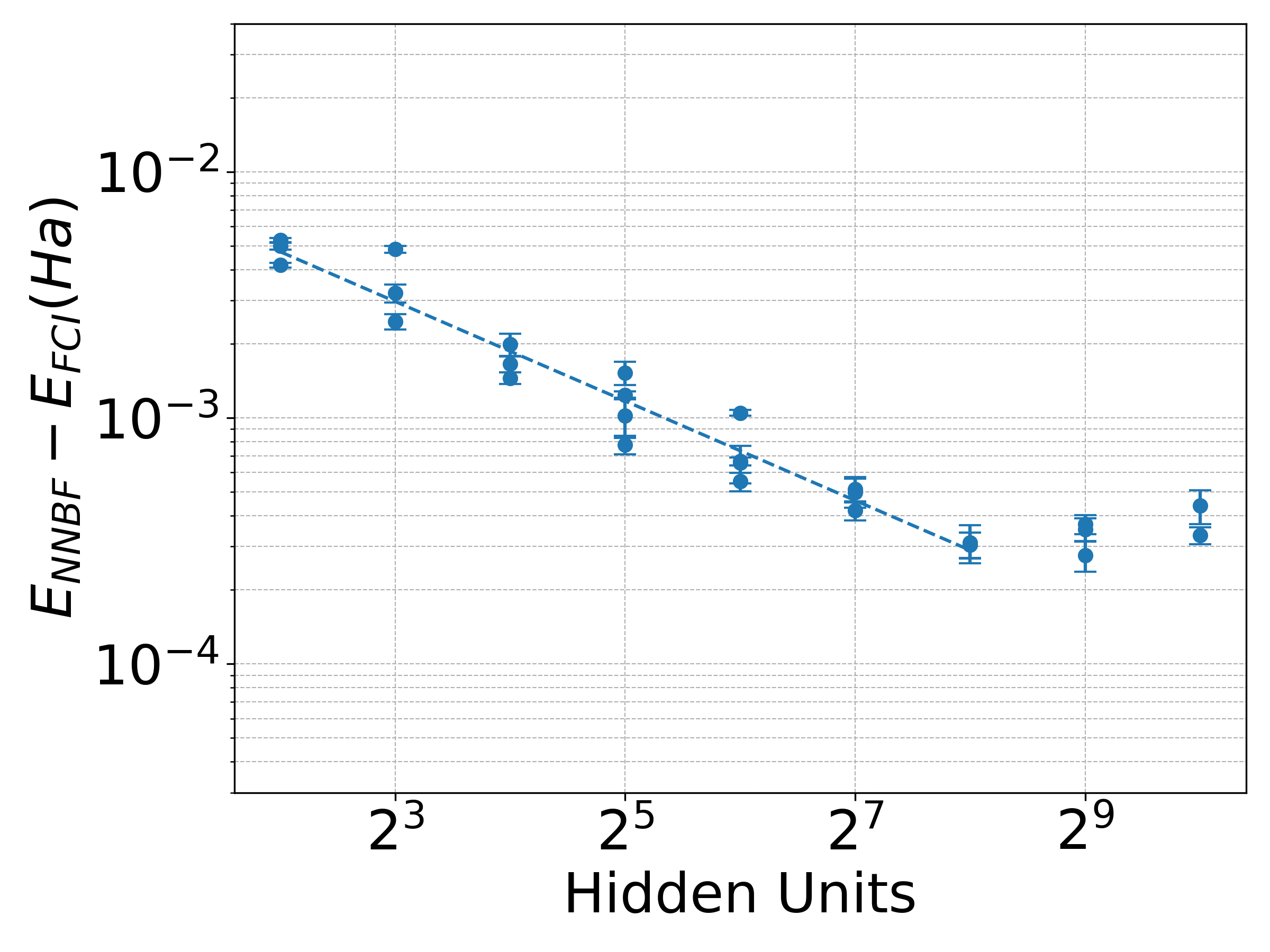}
    \caption{}
    \label{fig:NN_arch_limited_width}
  \end{subfigure}
  \hfill
  \begin{subfigure}{0.32\linewidth}
    \includegraphics[width=\linewidth]{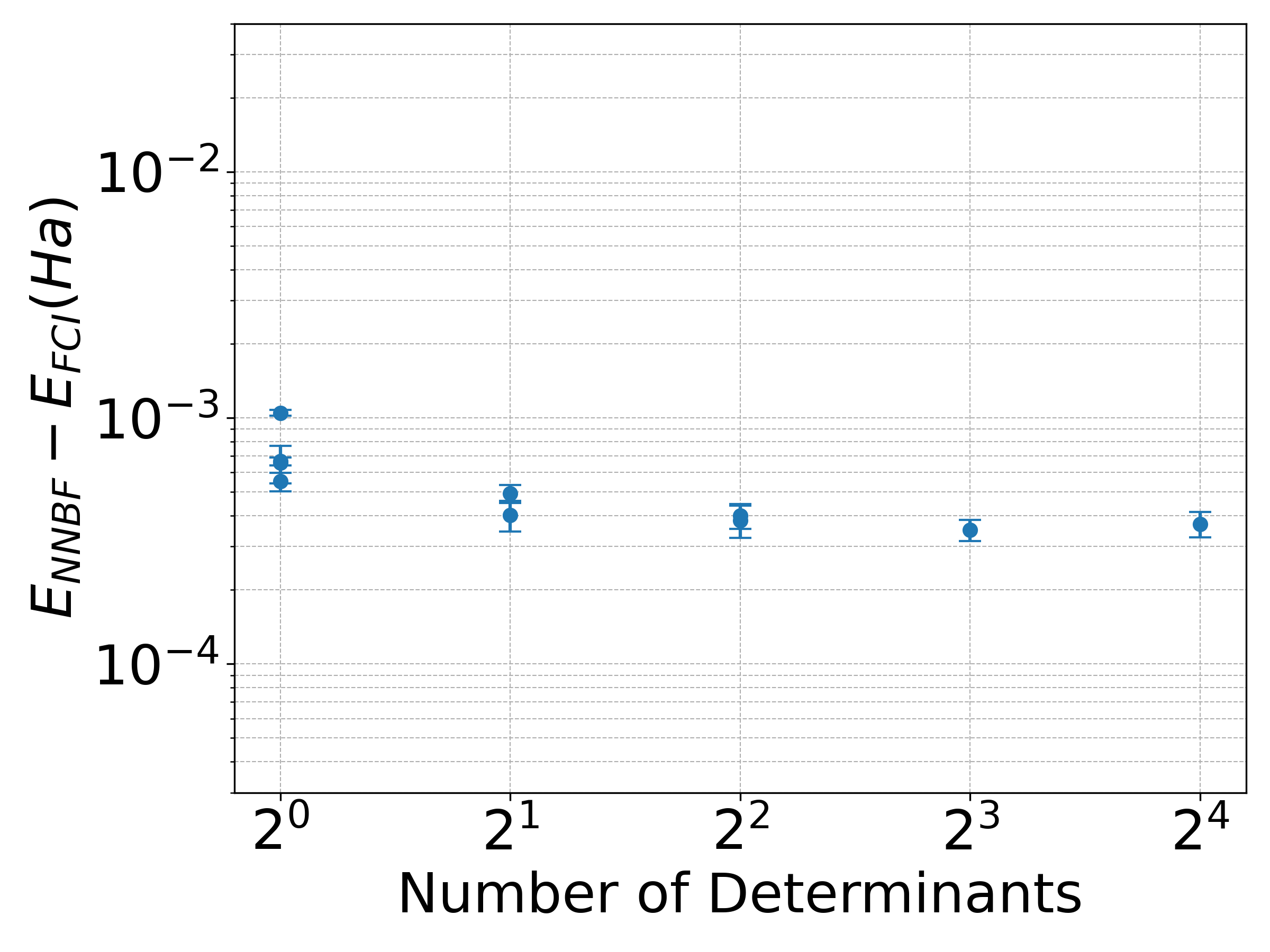}
    \caption{}
    \label{fig:NN_arch_limited_det}
  \end{subfigure}
\caption{Effects of network architecture on NNBF performance on the Li$_2$O with $N_u=8192$. (a) Effect of network depth. The improvement with more layers quickly saturates after 2 layers are added to NNBF states. Baselines from other NNQS works have been provided for comparison: QiankunNet \cite{Shang2023}, NAQS \cite{Barrett2022}, MADE \cite{Zhao2023}, QiankunNet$^*$ \cite{Wu2023}, RBM-SC \cite{Li2023}. The dark orange star denotes the best NNBF energy we have obtained with a greater $N_u$. (b) Effect of number of hidden units. Wider hidden layer continuously improves the accuracy with the absolute error drops at a rate of $\mathcal{O}(h^{-0.671855})$ with $R^2=0.952529$ until $h=256$. (c) Effect of number of determinants ($h=64$). Increasing the number of determinants reduces the energy error, but the improvement begins to plateau after reaching 4 determinants.}
\label{fig:NN_arch_limited}
\end{figure*}

It's intriguing to explore whether the impact of the network architecture on NNBF remains consistent when using the entire Hilbert space is not feasible in large systems, which is the typical regime of interest. Therefore, we conducted the same procedure as described in Section \ref{sec:NN_arch_full} on Li$_2$O with $N_u=8192$. The results are presented in Figure \ref{fig:NN_arch_limited}.

As the number of hidden units increases while maintaining 2 layers and 1 backflow determinant, accuracy uniformly improves and saturates after reaching 256 units. Additionally, adding more layers also increases accuracy, but the improvement from 2 to 3 layers is not significant, suggesting that additional layers may yield only minor gains. Expanding the number of determinants does reduce the energy error, with the improvement beginning to plateau after 4 determinants. The pattern shown in Figure \ref{fig:NN_arch_limited} is essentially consistent with that shown in Figure \ref{fig:NN_arch_full}.

A significant observation from Figure \ref{fig:NN_arch_limited_layer} is that our NNBF state with zero hidden layer, i.e. the configuration-dependent single particle orbitals are equal to  $\textbf{W} \textbf{x}_{input} + \textbf{b}$, yields a respectable energy of $-87.892004(67)$ Ha. Remarkably, this performance surpasses not only the CCSD baseline but also all existing NNQS results except QiankunNet \cite{Shang2023}.

\subsection{Interplay Between Batch Size And Network Architecture}

In the previous subsections, we considered the role of increasing the size of the neural network to improve the representability of the wave-function and hence its energy.  Somewhat surprisingly, we find that the final energy depends not only on the network architecture but also the batch size used in optimization.  In particular, in Figure \ref{fig:NN_arch_limited_layer}, we see that by using a larger batch size, we get a non-trivial improvement in the energy for Li$_2$O. In this section, we more systematically study the role of changing the batch size of the optimization for a fixed architecture. 

The primary role of the batch size $N_u$ is to determine the accuracy of the energy and gradients; notice that because we are using FSSC, larger batch sizes involve more unique independent configurations and therefore involve configurations that might not otherwise be seen.  

\begin{figure*} 
  \centering

  \begin{subfigure}{0.32\linewidth}
    \includegraphics[width=\linewidth]{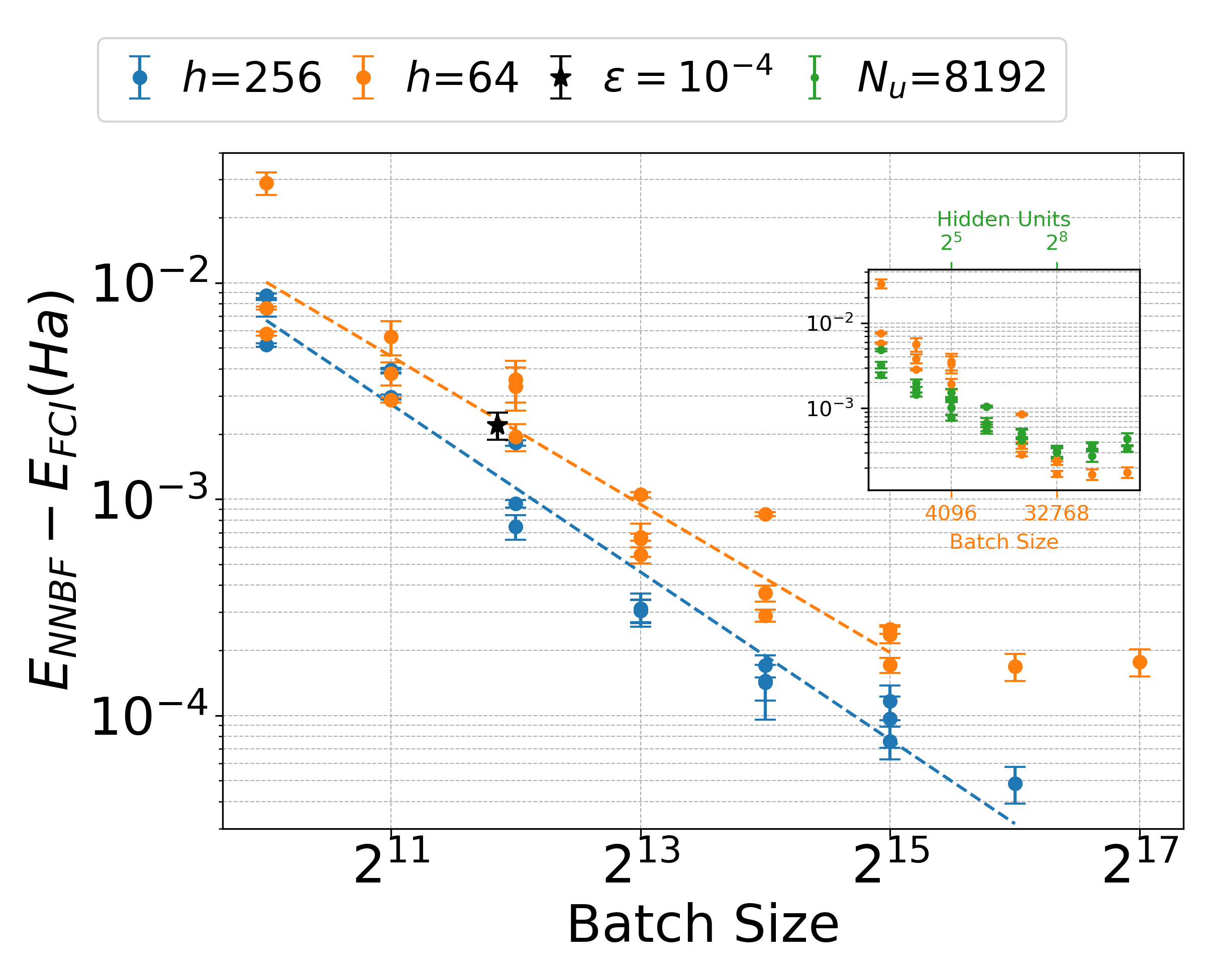}
    \caption{}
    \label{fig:NN_bs_Li2O}
  \end{subfigure}
  \hfill
  \begin{subfigure}{0.32\linewidth}
    \includegraphics[width=\linewidth]{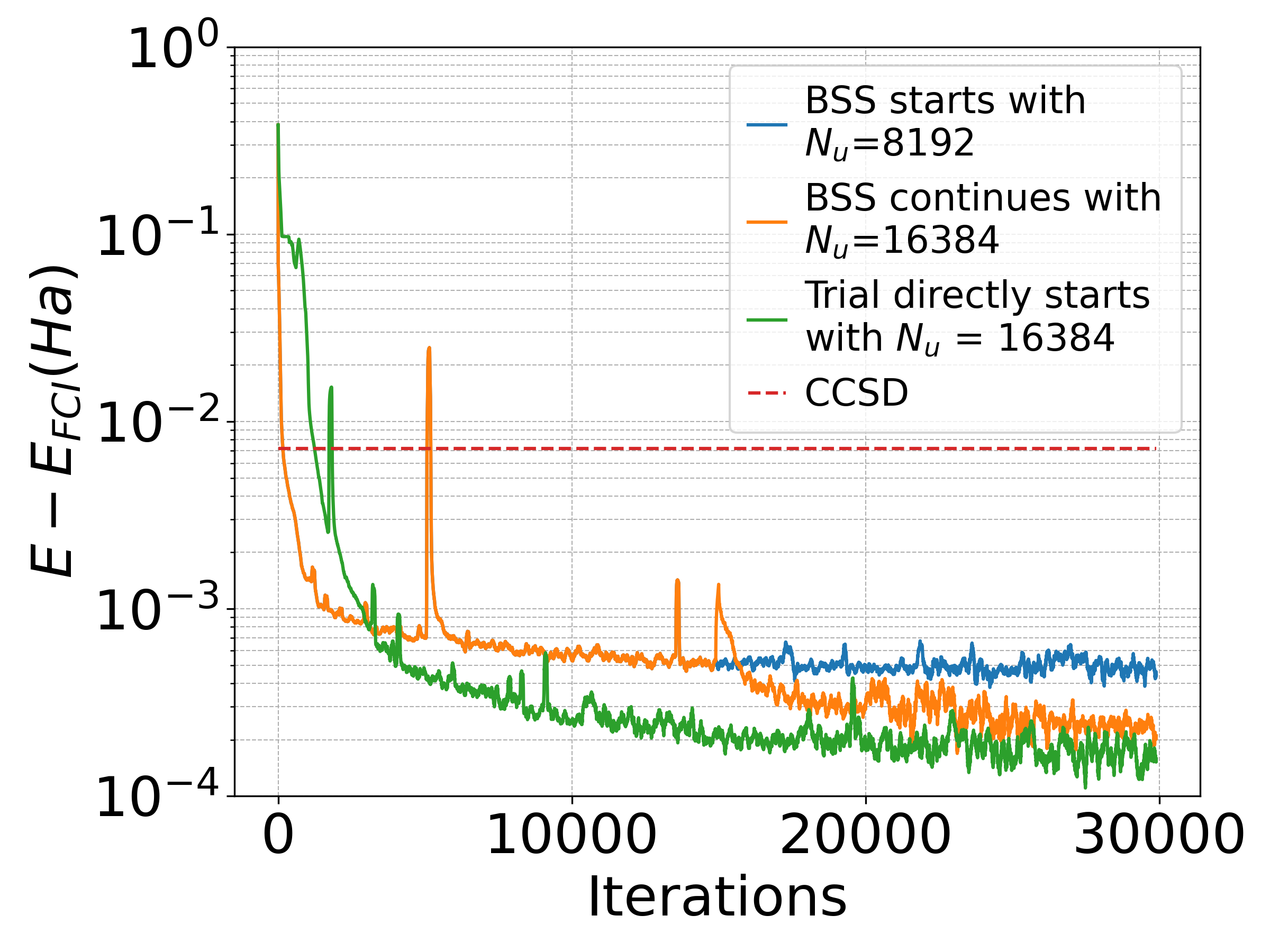}
    \caption{}
    \label{fig:BSS_Li2O}
  \end{subfigure}
  \hfill
  \begin{subfigure}{0.32\linewidth}
    \includegraphics[width=\linewidth]{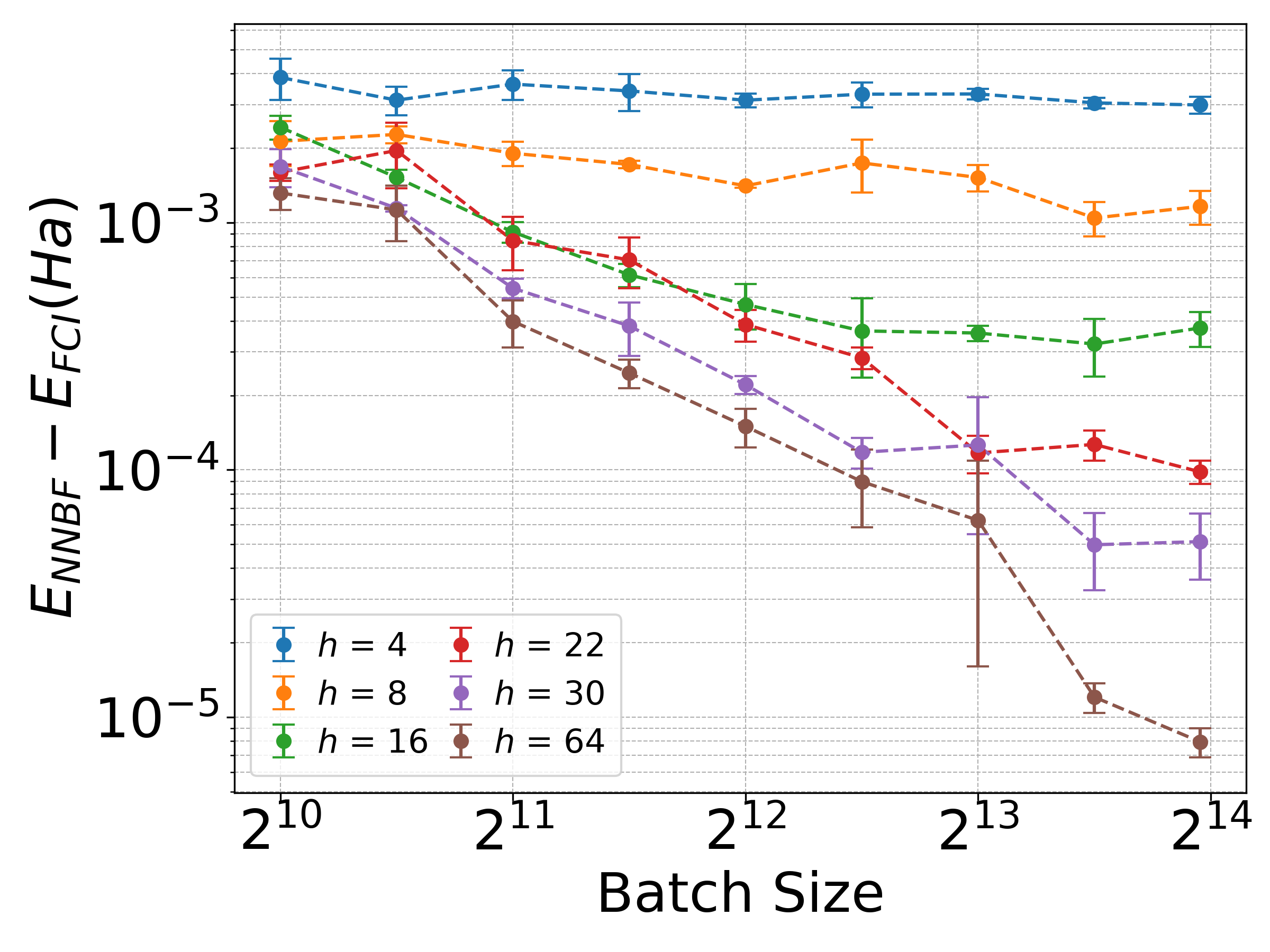}
    \caption{}        
    \label{fig:NN_bs_CH4}
  \end{subfigure}
\caption{(a) Experiments exploring the impact of batch size $N_u$ on energy improvements with $(L,h,D)=(2,64,1)$ and $(L,h,D)=(2,256,1)$ on lithium oxide. The energy error decreases approximately following $\mathcal{O}(N_u^{-1.137885})$ with an $R^2$ value of $0.900605$ for $h=64$ and $\mathcal{O}(N_u^{-1.288627})$ with $R^2=0.964273$ for $h=256$. Some data points from Figure \ref{fig:NN_arch_limited_width} are included for comparison in the inset. The black star represents a trial trained with $(L,h,D)=(2,64,1)$ using the sampling scheme from ref.~\onlinecite{Li2023}. Its x-position is determined by the average batch size at convergence. The energy closely aligns with the fitted line for the FSSC scheme, indicating that using a dynamic batch size does not offer a noticeable improvement in training. (b) Demonstration of the effectiveness of the batch size scheduling (BSS) strategy, with a moving average window of 100 applied for improved readability. (c) Experiments examining the dependence of energy improvements on batch size $N_u$ with various $h$ values on methane. Each data point represents the average and the standard deviation of the exact NNBF energy across 3 seeds.}
\label{fig:NN_bs}
\end{figure*}

To investigate $N_u$, we trained our model with batches of varying sizes for Li$_2$O, employing two sets of network architectures: an optimal one $(L,h,D)=(2,256,1)$ and a sub-optimal one $(L,h,D)=(2,64,1)$ identified in Section \ref{sec:NN_arch_limited}. Figure \ref{fig:NN_bs_Li2O} illustrates that increasing the batch size uniformly enhances the performance of the NNBF state for both network architectures. 
Remarkably, from the inset in Figure \ref{fig:NN_bs_Li2O}, we see that inferior architectures with larger batch size can reach lower energies than better architectures with smaller batch sizes. Both the improvement in energy from increasing the batch size and hidden units appears to scale polynomially (i.e. roughly linear on a log-log plot) with other hyperparameters fixed  exhibiting convergence rates of $\mathcal{O}(N_u^{-1.137885})$  and $\mathcal{O}(h^{-0.671855})$ respectively until a point of saturation.  Surprisingly increasing batch size converges faster; note that the computational complexity to increasing $N_u$  scales linearly while increasing $h$ scales quadratically  (since $L>1$; details are provided in Appendix \ref{sec:complexity}). 
Increasing $N_u$ also offers another inherent advantage: we can conveniently initiate our optimization at a smaller $N_u$, and once it has reached convergence, we can employ the trained NNBF state as the initial state for training with a larger batch size, as the network architecture remains unchanged. This enables a non-trivial portion of our optimization iterations to be conducted at smaller $N_u$, thereby saving computational time.
In Figure \ref{fig:BSS_Li2O}, we demonstrate both ramping up $N_u$ from a smaller $N_u$ as well as running the full optimization at a larger $N_u$.  We find that ramping up $N_u$ reaches very similar energies with the same number of iterations and has the additional benefit that, in the energy trace, there is less spikiness caused by selection of configurations from regions that haven't been well-trained. These spikes can significantly impair the initial training process, particularly in large systems, and potentially prevent the energy from descending. Initiating with a smaller $N_u$ can help mitigate this issue by exposing fewer new configurations to the algorithm at each iteration. This reduces the likelihood of encountering the not-well-trained regions before the NNBF state has generalized to unseen configurations. Notice that ramping $h$ is significantly more difficult as it requires some form of transfer learning to ``move'' the trained wave-function from a smaller to larger network. 

We further wish to understand if there is a relationship between the complexity of the network and the batch size required to saturate the optimization of that network.  
Another notable observation from Figure \ref{fig:NN_bs_Li2O} is that the energy for the network with $h=64$ begins to saturate within the range of batch sizes tested, while the network with $h=256$ shows ongoing improvements in energy. 
To better understand this, we focus on a smaller molecule, CH$_4$, where we can thoroughly investigate this correlation.
Figure \ref{fig:NN_bs_CH4} consistently shows that as the number of hidden units increases, the saturating batch size also grows larger. This further validates that a larger $N_u$ is required for a more expressive NNBF wave-function to reach the global minimum, mirroring a common relationship observed in traditional machine learning between neural network size and dataset size.

\section{Conclusions}
In this work, we show how to use NNBF with a deterministic FSSC optimization scheme to reach state-of-the-art energies on ab-initio second quantized molecular Hamiltonians.  On the large molecules, these energies are better than any other variational NNQS approaches as well as the CCSD method.  This gives strong evidence that NNBF is a powerful ansatz for second-quantized ab-initio QC. 
We additionally show that the FSSC scheme effectively captures crucial configurations, leading to faster, lower, and smoother energy convergence compared to conventional VMC with equivalent batch sizes.

We systematically investigated the impact of network architecture on NNBF performance. Increasing the number of hidden layers and backflow determinants reduces energy error but reaches saturation quickly.  The aspect of the network architecture which makes the largest difference in improving the energy is increasing the number of hidden units which seems to decrease the energy polynomially until saturation. These trends remain consistent across experiments both on small molecules optimized with batch sizes equal to the full Hilbert space as well as on larger molecules optimized with batch sizes much smaller than the full Hilbert space. 

Our analysis of batch size suggests that increasing it is crucial to improving the final optimized energy and, in fact, is a more efficient and effective strategy for improving NNBF energies compared to expanding hidden units in the regime where neither of these parameters is saturated.  Increasing the batch size is also less costly computationally and easier to ramp up as a function of optimization iteration. 

Further technical advancements, as explored in prior works \cite{Malyshev2023, Wu2023, Zhao2023}, hold the potential to further enhance the performance and efficiency of NNBF. Given the more favorable computational costs of our NNBF ansatz compared to RBMs and ARNs in the large basis set limit (see Appendix \ref{sec:complexity}), future research could focus on investigating how easily the NNBF model extrapolates to the complete basis set limit allowing closer comparison with continuum methods based on first quantization  \cite{Hermann2020, Pfau2020}.


\begin{acknowledgments}
We would like to express our gratitude for the insightful discussions with Zejun Liu and Kieran Loehr. This work utilized the Illinois Campus Cluster, a computing resource operated by the Illinois Campus Cluster Program in collaboration with the National Center for Supercomputing Applications and supported by funds from the University of Illinois at Urbana-Champaign. We also acknowledge support from the NSF Quantum Leap Challenge Institute for Hybrid Quantum Architectures and Networks (NSF Award 2016136).
\end{acknowledgments}


\appendix
\section{Experimental Setup}
\label{sec:experimental_setup}

In this work, we utilize the Adam optimizer to minimize the energy expectation value of our NNBF wave-function to approximate the ground state of various molecules with the STO-3G basis set. The default hyperparameters are shown in Table \ref{tab:default_hyperparameter}, and the energies for HF, CCSD, CCSD(T), and FCI baselines are calculated using the \textit{PySCF} software package \cite{pyscf}.  Following the training process with FSSC scheme whose detailed implementation is provided in Sec \ref{sec:fssc_implemenation}, we conduct a separate MCMC inference procedure to obtain the stochastic estimation of the energy of the trained NNBF state and report these numbers in all the experiments, except
for the demonstration of optimization progress with FSSC scheme. 

Sampling from the current unnormalized probability distribution $\Bar{p}_\theta(\bm{x})=\psi_\theta(\bm{x})^2$, each walker generates a Markov chain $\{\ket{\bm{x}_1} \rightarrow \ket{\bm{x}_2} \rightarrow \ket{\bm{x}_3} \rightarrow \cdots \}$ using the Metropolis–Hastings algorithm, proposing a swap between an occupied site and an unoccupied one at each iteration. Subsequently, samples are downsampled from the Markov chain at an interval $K_1$, forming $\{\ket{\bm{x}_1}, \ket{\bm{x}_K}, \ket{\bm{x}_{2K}} , \cdots, \ket{\bm{x}_{MK}}\}$. In this work, we typically set $K_1=10N_e$ and $M=1000$, with $N_w=1024$ walkers operating concurrently. The $M$ samples will be averaged within each walker to generate $N_w$ energies. Subsequently, the energy expectation value is computed by averaging over these energies, while the uncertainties are derived from the variance among these $N_w$ energies to mitigate the influence of auto-correlation time. The NNBF energy and error bars presented in this report are obtained through this procedure. Note that we followed the approach outlined in ref.~\onlinecite{ForemanMackey2013} for initializing walkers. Specifically, we selected the $M=8$ most dominant configurations from the last core space utilized in the FSSC scheme. These configurations were used to set the initial positions of the $N_w$ walkers based on their relative probabilities, and the walkers underwent $K_2=200$ burn-in steps before starting to collect samples. This approach helps prevent the walkers from becoming trapped in low-probability modes of a multi-modal probability landscape, facilitating faster convergence to equilibrium.

The same sampling routine can be employed to draw MCMC samples during the training phase with MCMC scheme, with the distinction that the network is updated at an interval $K$, and a larger $N_w$ may be utilized to ensure an accurate stochastic estimate of energy and its gradient.

\begin{table}[]
\centering
\begin{tabular}{cc}
    \toprule
    Hyperparameter & Value \\
    \midrule
    Energy unit & Hartree \\
    Atomic orbital & STO-3G \\
    Molecular orbital & canonical Hartree-Fock \\
    Framework & JAX \\
    Precision & float32 \\
    Optimizer & Adam \\
    Adam's $\beta_1$ & 0.9 \\
    Adam's $\beta_2$ & 0.999 \\
    Adam's $\epsilon$ & $1\times10^{-8}$ \\
    Learning rate & $10^{-3}\times(1+10^{-4}t)^{-1}$ \\
    Training iterations & 50000 \\
    Number of pretraining walkers & 8192 \\
    Pretraining iterations & 500 \\
    Number of dominant configurations  & 8 \\
    for walker initialization & \\
    MCMC burn-in steps & 200 \\
    right after initialization & \\
    Number of MCMC walkers & 1024 \\
    MCMC downsample interval & $10N_e$ \\
    MCMC iterations & 1000 \\
    Number of hidden layers  & 2 \\
    Hidden units & 256 \\
    Number of determinants  & 1\\
    \bottomrule
\end{tabular}
\caption{Default hyperparameters used for all experiments in the paper, unless explicitly stated otherwise}
\label{tab:default_hyperparameter}
\end{table}

\begin{table}[]
\centering
\begin{tabular}{cc}
    \toprule
    symbol & description \\
    \midrule
    $N_e$ & number of electrons \\
    $N_o$ & number of spin-orbitals \\ 
    $N_t$ & number of physically valid configurations \\
    $N_u$ & number of unique configurations \\
    $L$ & number of hidden layers \\
    $h$ & hidden units \\
    $D$ & number of determinants \\
    $N_w$ & number of MCMC walkers \\
    \bottomrule
\end{tabular}
\caption{Table of notations}
\label{tab:notation}
\end{table}


\section{CISD Pre-training}
\label{sec:CISD_pretraining}

Machine learning scientists understand that initializing the model close to the global minimum significantly aids training. This insight prompted us to pretrain the network to approximate the configuration interaction with single and double excitations (CISD) state, computed using \textit{PySCF} \cite{pyscf}, before optimizing for the expected energy. The pretraining loss function we employed is the negative logarithm of the fidelity between the current NNBF ansatz and the CISD wave-function:
\begin{equation}\label{eq:fidelity}
\mathcal{F}(\theta)=-\ln{\frac{\abs{\braket{\Psi_\theta}{\Psi_T}}^2}{\braket{\Psi_\theta}{\Psi_\theta} \braket{\Psi_T}{\Psi_T}}}
\end{equation} 
where each component can be formulated as expectation values such as:
\begin{equation}
\braket{\Psi_\theta}{\Psi_T} = \mathbb{E}_{p_{\theta}(\bm{x})}\left[ \frac{\braket{\bm{x}}{\Psi_T}}{\braket{\bm{x}}{\Psi_\theta}} \right]
\end{equation}
Also, the gradient of $ \mathcal{F}(\theta)$ can be expressed as:
\begin{align}\label{eq:fidelity_grad}
\frac{\partial\mathcal{F}}{\partial\theta}=-2\Re\left\{\mathbb{E}_{p_{\theta}(\bm{x})}\left[ \left( \frac{1}{\braket{\Psi_\theta}{\Psi_T}} \frac{\braket{\bm{x}}{\Psi_T}}{\braket{\bm{x}}{\Psi_\theta}} \right.\right.\right. \nonumber \\
\left.\left.\left. -\frac{1}{\braket{\Psi_\theta}{\Psi_\theta}}  \right) \nabla_\theta \ln{\braket{\Psi_\theta}{\bm{x}}} \right]\right\}.
\end{align} Here, ~\eqref{eq:fidelity} and ~\eqref{eq:fidelity_grad} are estimated using the standard MCMC procedure, where the samples are drawn from the unnormalized probability $\Bar{p}_\theta(x)=\psi_\theta(x)^2$.

\begin{figure}
\centering
\includegraphics[width=\linewidth]{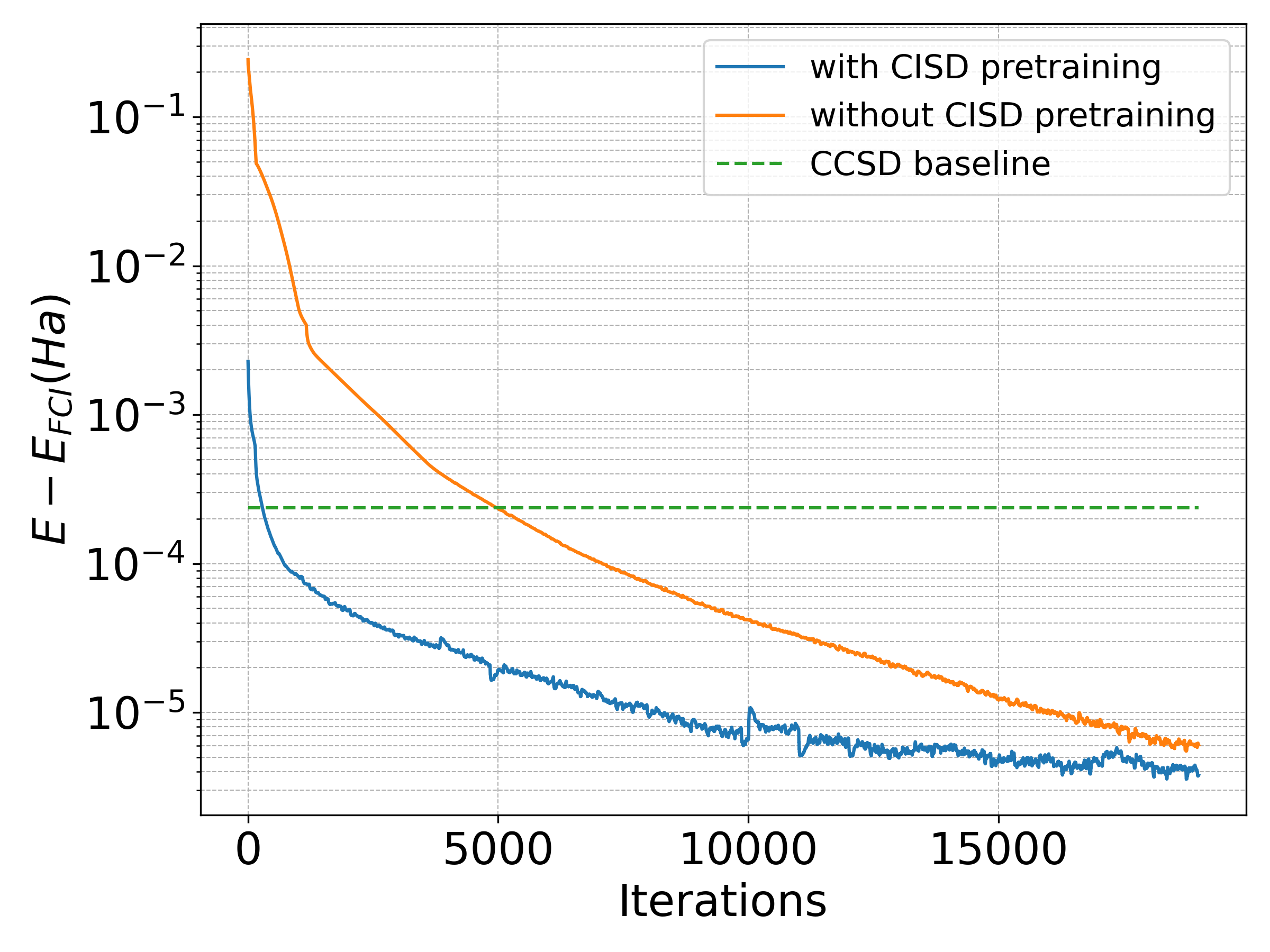}
\caption{Optimization progress for CH$_4$, with and without CISD pretraining ($L=1$, $h=512$, $N_u=N_t$). The training iteration comprises 20000 steps, and a moving average window of 1000 is applied for better readability.}
\label{fig:CISD_demo}
\end{figure}

The comparison between the trials with and without pretraining is depicted in Figure \ref{fig:CISD_demo}. It's evident that the objective energy converges faster with CISD pretraining, as it avoids optimizing through a region known to be physically uninteresting. Using MCMC instead of FSSC to draw samples could prevent the time-consuming local energy calculation. Hence, prepending the pretraining to our training routine seems to be a good practice, although we did observe that pretraining stranded the neural network in a poor local optimum in rare cases.

\section{Computational Complexity Analysis}\label{sec:complexity}

Using the notations in Table \ref{tab:notation}, we can express the computational cost of NNBF as follows: evaluating the MLP scales as $\mathcal{O}(N_o h + (L-1)h^2 + hDN_oN_e)$, while evaluating the determinants scales as $\mathcal{O}(DN_e^3)$. This yields a total computational cost of $\mathcal{O}(N_o h + (L-1)h^2 + hDN_oN_e + DN_e^3)$ for calculating the amplitude for one configuration and $\mathcal{O}(N_u[N_o h + (L-1)h^2 + hDN_oN_e + DN_e^3])$ for a batch of $N_u$ configurations.

In the large basis limit, where $N_o$ increases while $N_e$ is fixed, the time complexity (without optimization) of NNBF would be $\mathcal{O}(N_o)$ if we also fix the network architecture $(L,h,D)$. In comparison, for a simple shallow RBM, its computational cost is $\mathcal{O}(N_o + \alpha N_o + \alpha N_o^2) = \mathcal{O}(N_o^2)$, where $\alpha$ is the hidden layer density. Similarly, for an ARN, the computational cost for calculating the conditional probability distribution $p_\theta(x_i | x_1,x_2,\dots,x_{i-1})$ is at least $\Omega(i)$ (transformer-based model has even worse complexity, $\mathcal{O}(i^2)$), resulting in a cost of $\Omega(N_o^2)$ for exact sampling of one unique configuration, $p_\theta(\bm{x}) = \prod_{i=2}^{N_o} p_\theta(x_i | x_1,x_2,\dots,x_{i-1})$. Both architectures exhibit a computational cost quadratic in $N_o$, not $N_e$. This distinction renders our NNBF model computationally more favorable when $N_o$ is large with $N_e$ fixed.


\clearpage 
\bibliography{reference}

\end{document}